\documentclass[aip, apl, preprint, amsmath, graphicx]{revtex4-1}

\usepackage{graphics}
\usepackage{graphicx}
\usepackage{bm}
\usepackage{color}

\begin{document}


\title{Interface-induced anomalous Nernst effect in Fe$_3$O$_4$/Pt-based heterostructures}



\author{R. Ramos}
\email[]{ramosr@imr.tohoku.ac.jp}
\affiliation{Advanced Institute for Materials Research, Tohoku University, Sendai 980-8577, Japan}

\author{T. Kikkawa}
\affiliation{Advanced Institute for Materials Research, Tohoku University, Sendai 980-8577, Japan}
\affiliation{Institute for Materials Research, Tohoku University, Sendai 980-8577, Japan}

\author{A. Anad\'{o}n}
\affiliation{IMDEA Nanociencia, Campus de Cantoblanco, 28049 Madrid, Spain}

\author{I. Lucas}
\affiliation{Instituto de Nanociencia de Arag\'{o}n, Universidad de Zaragoza, E-50018 Zaragoza, Spain}
\affiliation{Departamento de F\'{i}sica de la Materia Condensada, Universidad de Zaragoza, E-50009 Zaragoza, Spain}
\affiliation{Fundaci\'{o}n Instituto de Nanociencia de Arag\'{o}n, E-50018 Zaragoza, Spain}

\author{T. Niizeki}
\affiliation{Advanced Institute for Materials Research, Tohoku University, Sendai 980-8577, Japan}

\author{K. Uchida}
\affiliation{Institute for Materials Research, Tohoku University, Sendai 980-8577, Japan}
\affiliation{National Institute for Materials Science, Tsukuba 305-0047, Japan}
\affiliation{Center for Spintronics Research Network, Tohoku University, Sendai 980-8577, Japan}

\author{P. A. Algarabel}
\affiliation{Departamento de F\'{i}sica de la Materia Condensada, Universidad de Zaragoza, E-50009 Zaragoza, Spain}
\affiliation{Instituto de Ciencia de Materiales de Arag\'{o}n, Universidad de Zaragoza and Consejo Superior de Investigaciones Cient\'{i}ficas, 50009 Zaragoza, Spain}

\author{L. Morell\'{o}n}
\affiliation{Instituto de Nanociencia de Arag\'{o}n, Universidad de Zaragoza, E-50018 Zaragoza, Spain}
\affiliation{Departamento de F\'{i}sica de la Materia Condensada, Universidad de Zaragoza, E-50009 Zaragoza, Spain}
\affiliation{Fundaci\'{o}n Instituto de Nanociencia de Arag\'{o}n, E-50018 Zaragoza, Spain}

\author{M. H. Aguirre}
\affiliation{Instituto de Nanociencia de Arag\'{o}n, Universidad de Zaragoza, E-50018 Zaragoza, Spain}
\affiliation{Departamento de F\'{i}sica de la Materia Condensada, Universidad de Zaragoza, E-50009 Zaragoza, Spain}
\affiliation{Fundaci\'{o}n Instituto de Nanociencia de Arag\'{o}n, E-50018 Zaragoza, Spain}
\affiliation{Laboratorio de Microscop\'{i}as Avanzadas, Universidad de Zaragoza, E-50018 Zaragoza, Spain}

\author{M. R. Ibarra}
\affiliation{Instituto de Nanociencia de Arag\'{o}n, Universidad de Zaragoza, E-50018 Zaragoza, Spain}
\affiliation{Departamento de F\'{i}sica de la Materia Condensada, Universidad de Zaragoza, E-50009 Zaragoza, Spain}
\affiliation{Fundaci\'{o}n Instituto de Nanociencia de Arag\'{o}n, E-50018 Zaragoza, Spain}
\affiliation{Laboratorio de Microscop\'{i}as Avanzadas, Universidad de Zaragoza, E-50018 Zaragoza, Spain}

\author{E. Saitoh}
\affiliation{Advanced Institute for Materials Research, Tohoku University, Sendai 980-8577, Japan}
\affiliation{Institute for Materials Research, Tohoku University, Sendai 980-8577, Japan}
\affiliation{Center for Spintronics Research Network, Tohoku University, Sendai 980-8577, Japan}
\affiliation{Department of Applied Physics, The University of Tokyo, Tokyo 113-8656, Japan}
\affiliation{Advanced Science Research Center, Japan Atomic Energy Agency, Tokai 319-1195, Japan}


\date{\today}

\begin{abstract}
We have studied the anomalous Nernst effect (ANE) in [Fe$_3$O$_4$/Pt]-based heterostructures, by measuring the ANE-induced electric field with a magnetic field applied normal to the sample surface, in the perpendicular magnetized configuration, where only the ANE is expected. An ANE voltage is observed for [Fe$_3$O$_4$/Pt]$_n$ multilayers, and we further investigated its origin by performing measurements in [Fe$_3$O$_4$/Pt/Fe$_3$O$_4$] trilayers as a function of the Pt thickness. Our results suggest the presence of an interface-induced ANE. Despite of this ANE, the spin Seebeck effect is the dominant mechanism for the transverse thermoelectric voltage in the in-plane magnetized configuration, accounting for about 70 \% of the measured voltage in the multilayers. 
\\
\end{abstract}


\maketitle
\begin{figure}[htp!] \center
    \includegraphics[width=8.5cm, keepaspectratio]{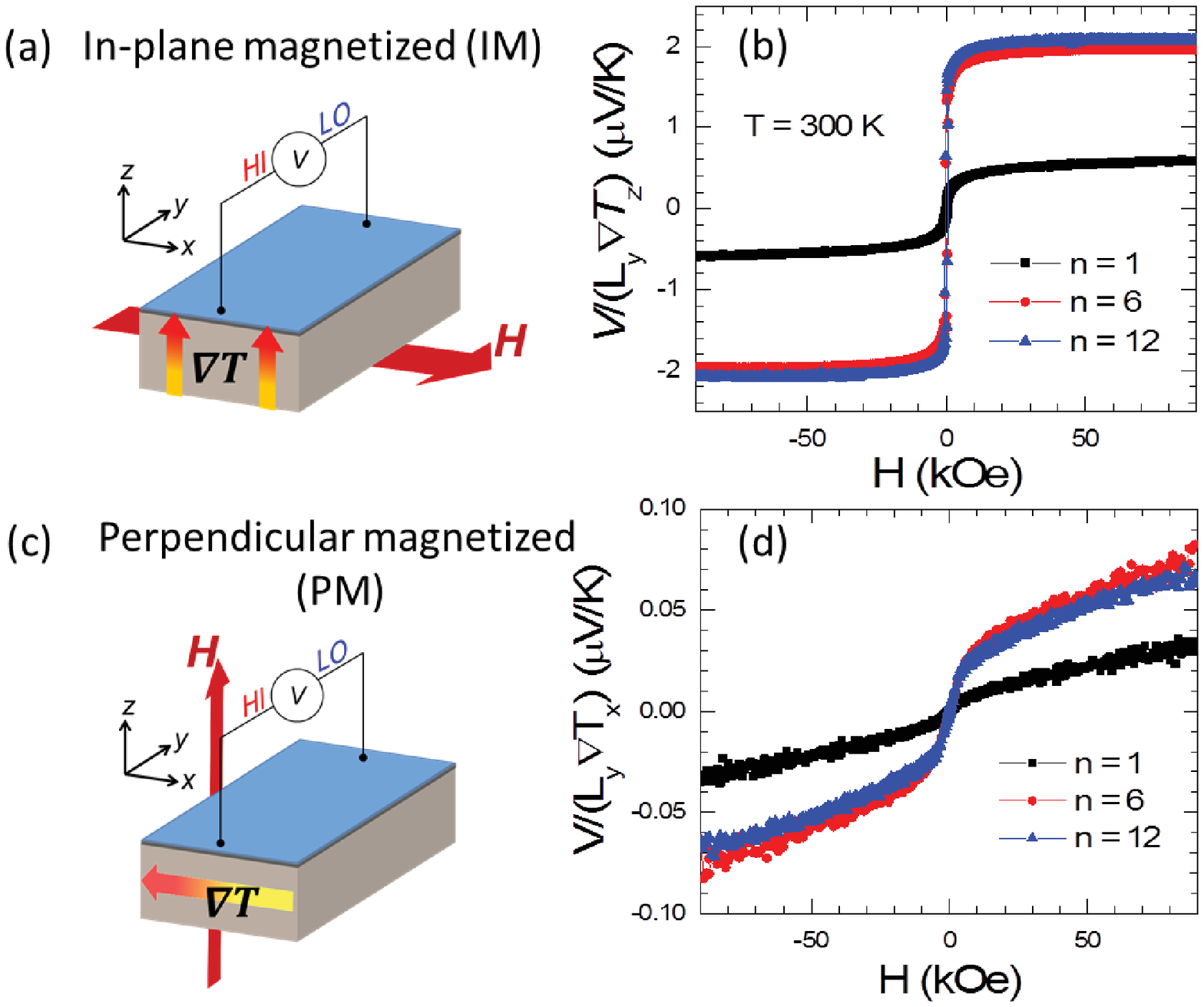}
  \caption{(Color online) (a) Schematic illustration of the in-plane magnetized (IM) configuration conventionally used for the measurement of the longitudinal spin Seebeck effect (SSE) in magnetic insulators. (b) Thermoelectric voltage (SSE + ANE) measured in [Fe$_3$O$_4$(23)/Pt(7)]$_\text{n}$ (thickness in nm) multilayers in the IM configuration.  (c) Schematic illustration of the perpendicular magnetized (PM) configuration for the measurement of the anomalous Nernst effect (ANE). (d) Thermoelectric voltage (ANE) measured in the PM configuration for [Fe$_3$O$_4$(23)/Pt(7)]$_\text{n}$ multilayers.}
  \label{fig1}
\end{figure}

\begin{figure*}[htp!] \center
    \includegraphics[width=17cm, keepaspectratio]{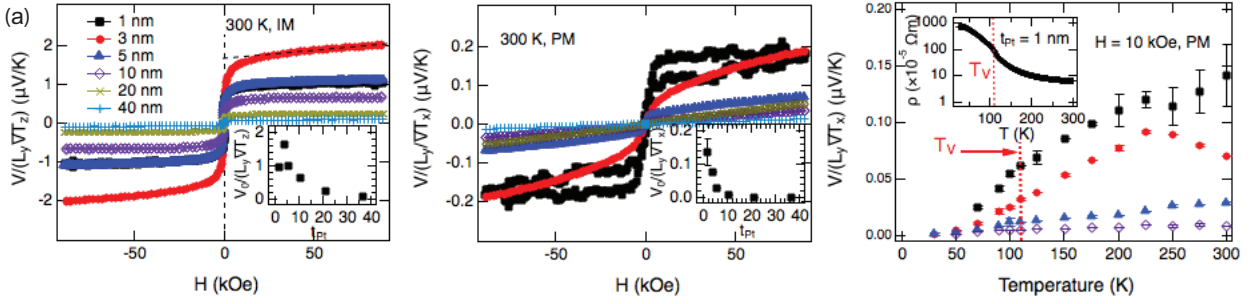}
  \caption{(Color online) (a,b) Magnetic field dependence of the transverse thermoelectric voltage measured in the (a) IM and (b) PM configurations for [Fe$_3$O$_4$(50)/Pt($t_\text{Pt}$)/Fe$_3$O$_4$(50)] trilayers with different nominal Pt thicknesses. Inset of (a), (b) show the magnitude of the intercept at zero field of the measured voltage [$V_0/(L_y\nabla T_{(z,x)})$], determined by linear extrapolation of the high field data as schematically depicted by the dashed line of (a). (c) Temperature dependence of the magnitude of the transverse thermoelectric voltage in the PM configuration at $H =$ 10 kOe for $t_\text{Pt}$ between 1 and 10 nm, obtained from the measured magnetic field dependence of the voltage at each temperature. Inset shows the temperature dependence of the resistivity of a Fe$_3$O$_4$/Pt(1)/Fe$_3$O$_4$ trilayer, showing that for this $t_\text{Pt}$ value the Fe$_3$O$_4$ layer dominates the transport properties.}
  \label{fig2}
\end{figure*}

\begin{figure}[htp!] \center
    \includegraphics[width=8.5cm,keepaspectratio]{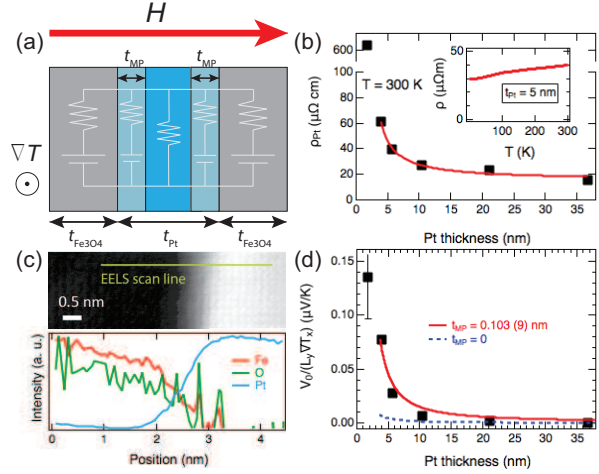}
  \caption{(Color online) (a) Schematic diagram of the model used to include a magnetic layer region between the non-magnetic Pt and the Fe$_3$O$_4$ layer (b) Thickness dependence of the Pt resistivity and fit using conventional transport  formula for metals [Pt thickness ($t_\text{Pt}$) measured by X-ray reflectivity is shown]. (c) STEM image of the Fe$_3$O$_4$/Pt interface of a [Fe$_3$O$_4$(50)/Pt(10)/Fe$_3$O$_4$(50)] trilayer (top) and elemental maps measured by EELS (Fe and O) and EDX (Pt) at the interface in the region indicated by the scan line (bottom).\cite{Kikkawa2017} (d) Thickness dependence of the ANE in the trilayers and fit (red line) of the experimental data using Eq. \ref{eq:E_MP_thickness_dep} in the range of Pt thickness from 3 to 40 nm (blue dashed line shows the electric field due to ANE of Fe$_3$O$_4$ layer only, $t_\text{MP}$ = 0).}
  \label{fig3}
\end{figure}

\begin{figure}[htp!] \center
    \includegraphics[width=8.5cm,keepaspectratio]{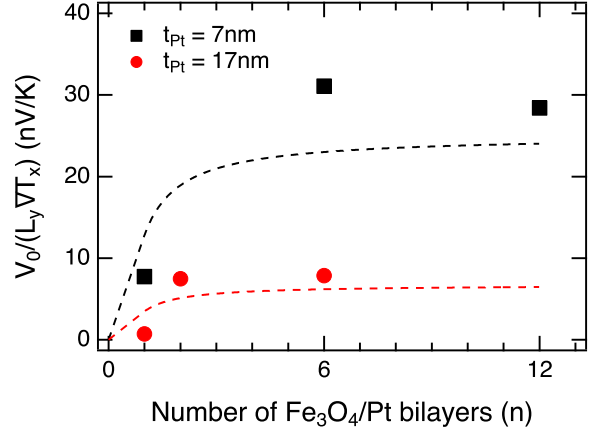}
  \caption{(Color online) Magnitude of the ANE obtained from the zero field intercept in the PM configuration for [Fe$_3$O$_4$(23)/Pt(7)]$_\text{n}$ (black squares) and [Fe$_3$O$_4$(34)/Pt(17)]$_\text{n}$ (red circles) multilayers. The dashed lines show the ANE contribution estimated using Eq. \ref{eq:E_MP_ML}, considering $t_\text{MP}$= 0.1 nm.}
  \label{fig4}
\end{figure}

 
Thermoelectricity deals with the study of heat-to-electricity interconversion processes, having the potential for the development of waste heat energy-harvesting technologies. Since the discovery of the spin Seebeck effect (SSE) \cite{uchida:nat2008, uchida:sse-insulator} a new heat-to-electricity  conversion paradigm in magnetic systems was established and has been extensively studied in a wide range of materials.\cite{Uchida2014a, Uchida2016} This invigorated the field of spin caloritronics,\cite{bauer:spinCaloritronics, Boona2014} which studies the interaction between heat, electron and spin currents.  In the SSE, a spin current\cite{Maekawa2013} is generally created in a ferromagnetic material (F) upon application of a temperature gradient, and electrically detected in an adjacent normal metal (N) by the inverse spin Hall effect (ISHE).\cite{Saitoh2006a}\\
The presence of the SSE in magnetic insulators,\cite{uchida:sse-insulator} has potential advantages over conventional thermoelectrics due to lower heat dissipation losses. Its experimental geometry, with the thermal and electric current paths perpendicular to each other, is also advantageous for the implementation of thin film and flexible thermoelectric devices.\cite{Kirihara2012, Kirihara2016} However, the low magnitude of the measured voltage is an obstacle for the development of potential SSE applications. In this regard, different possibilities are currently being explored,\cite{Uchida2016} such as increasing the spin current detection efficiency,\cite{Niimi2012, jiang2016} spin Hall thermopiles,\cite{Uchida2012d, Ramos2016} [F/N]$_n$ multilayers\cite{Ramos2015, Ramos2016, Ramos2017, Lee2015a, Uchida2015a, Shiomi2015, Uchida2017, Ramos2018} or bulk composite systems.\cite{Boona2016}\\ 
The ISHE-induced electric field, driven by the thermally generated spin currents in the SSE, can be written as:
 \begin{equation}
\label{eq:Eishe}
  \mathbf{{E}}_{\text{ISHE}}=\theta_{\text{SH}}\rho(\mathbf{J}_{\text{S}} \times \bm{\sigma}),
\end{equation}
where $\theta_{\text{SH}}$ and $\rho$ denote the spin Hall angle and electric resistivity of N, respectively. $\mathbf{E}_{\text{ISHE}}$, $\mathbf{J}_{\text{S}}$, and $\bm{\sigma}$ are the ISHE generated electric field, spatial direction of spin current (perpendicular to the F/N interface) and spin-polarization vector (parallel to the magnetization, $\mathbf{M}$). The geometry for the detection of the ISHE is similar to that of the ANE in electrically conductive ferromagnets, which is phenomenologically described by the expression:
 \begin{equation}
\label{eq:Eane}
  \mathbf{{E}}_{\text{ANE}}= Q_\text{S}\mu_0(\nabla T \times \mathbf{M}),
\end{equation}
where  $Q_\text{S}$, $\mu_0$, $\nabla T$, and $\mathbf{E}_{\text{ANE}}$ are ANE coefficient, the vacuum permeability, applied thermal gradient and ANE-induced electric field, respectively. Therefore, when using ferromagnetic metals, care must be taken in order to separate the contribution of the SSE from that of the ANE.\cite{Wu2017} In the case of Fe$_3$O$_4$/Pt the ANE contribution from Fe$_3$O$_4$ is expected to be negligibly small due to the resistivity of Fe$_3$O$_4$ being two orders of magnitude larger than that of Pt.\cite{Ramos2013}\\
Another source of ANE-driven electric field can be due to a possible magnetic proximity effect (MPE) at the Fe$_3$O$_4$/Pt interface. In the case of YIG/Pt and other insulating ferrimagnetic oxides the MPE has been shown to be negligibly small.\cite{geprags2012, Kuschel2015, Valvidares2016, Collet2017} However, recent X-ray magnetic circular dichroism (XMCD) measurements in a Fe$_3$O$_4$/Pt/Fe$_3$O$_4$ trilayer,\cite{Kikkawa2017} show a large induced magnetic moment in the Pt interlayer, therefore investigating its effect on the thermally-driven magnetotransport properties\cite{Bougiatioti2017} is important to gain further insight of the heat-to-electricity conversion process in [Fe$_3$O$_4$/Pt]-based heterostructures. To this purpose, we have performed measurements of the longitudinal SSE,\cite{Uchida2010i} with an in-plane magnetic field (IM) [Fig. \ref{fig1}(a)], and the ANE with a magnetic field applied perpendicular to the sample surface (PM) [Fig. \ref{fig1}(c)]. The measurements in the PM configuration allow to unambiguously determine the ANE due to the fact that, in this configuration, any injected spin current is parallel to the direction of the magnetic field $\mathbf{H}$ ($\mathbf{J}_{\text{S}} \parallel\mathbf{H} \parallel z$) and $E_\text{ISHE}$ = 0 due to the ISHE geometry  (Eq. \ref{eq:Eishe}).\cite{Kikkawa2013d, Uchida2014a}\\
Here, we have studied two type of Fe$_3$O$_4$/Pt-based heterostructures: [Fe$_3$O$_4$/Pt]$_n$ multilayers (these are the same samples used in our previous studies),\cite{Ramos2015, Ramos2016, Ramos2017, Uchida2017, Ramos2018} and [Fe$_3$O$_4$/Pt($t_\text{Pt}$)/Fe$_3$O$_4$] trilayers with nominal Pt thicknesses ranging from 1 to 40 nm. The Fe$_3$O$_4$ films were grown on MgO(001) substrates by pulsed laser deposition in an ultrahigh-vacuum chamber (UHV). The Pt films were sequentially deposited by DC magnetron sputtering in the same UHV chamber. The substrate temperature during the entire thin film growth process was kept at 480 $^\circ$C. The structural quality of the samples was confirmed by x-ray diffraction and high angle annular dark field (HAADF) scanning transmission electron microscopy (STEM), detailed information about sample preparation and characterization can be found elsewhere.\cite{Ramos2015} For the thermoelectric measurements we apply a constant heat power, inducing a thermal gradient ($\nabla T$) in the $z$ direction. A magnetic field ($\bf{H}$) with magnitude $H$ is swept in the $x$ direction while the SSE voltage in the Pt film is measured along the $y$ direction ($V$) [Fig. \ref{fig1}(a)].\cite{Uchida2010i, Uchida2014a, Uchida2016} For the PM measurements, $\nabla T$ and $\bf{H}$ are applied parallel to the sample $x$ and $z$ directions, respectively [Fig. \ref{fig1}(c)]. The sample dimensions are $L_\text{x}$ = 2 mm, $L_\text{y}$ = 7 mm and $L_\text{z}$ = 0.5 mm.\\
Figures \ref{fig1}(b) and (d) show the results of the transverse thermoelectric voltage for the [Fe$_3$O$_4$/Pt]$_\text{n}$ multilayers in the IM and PM configurations, respectively. The voltage measured in the IM configuration is more than one order of magnitude larger than that observed in the PM configuration. However, to quantitatively compare these voltages, we need to know the temperature gradient distribution in the sample. In the PM configuration both multilayer and substrate are subjected to the same thermal gradient, since they are in direct contact to the AlN plates.  However, in the IM configuration we need to evaluate the thermal gradient distribution across the thickness direction ($\parallel z$). By considering heat conservation across the direction of applied heat current: $\kappa_\text{MgO}\nabla T_\text{MgO}$ = $\kappa_\text{ML}\nabla T_\text{ML}$, and using the previously reported values for the thermal conductivities of MgO\cite{Dalton2013} and [Fe$_3$O$_4$/Pt]$_n$ multilayers,\cite{Ramos2015} we obtain $\nabla T_\text{ML}=\frac{\kappa_\text{MgO}}{\kappa_\text{ML}}\nabla T_\text{MgO} \sim 20 \nabla T_\text{MgO}$ for our samples.\cite{SupplMat} (Suppl. Mat.). Then, we can estimate that the ANE in the [Fe$_3$O$_4$/Pt]$_n$ multilayers accounts for about 30 \% of the measured voltage in the IM configuration, indicating that there is a non-negligible contribution from the spin-polarized conduction electrons to the transverse thermoelectric voltage, although significantly smaller than the SSE voltage.\\
In order to gain further insight into the nature of the ANE contribution, we systematically measured the transverse thermoelectric voltage in [Fe$_3$O$_4$/Pt($t_\text{Pt}$)/Fe$_3$O$_4$] trilayers with different platinum thicknesses, $t_\text{Pt}$. Figures \ref{fig2}(a) and (b) show the voltages measured in the IM and PM configurations, respectively. As shown in the inset of Fig. \ref{fig2}(a) the voltage in the IM configuration continuously increases as the Pt thickness decreases with a maximum at 3 nm and then rapidly decreasing, in agreement with previously reported Pt thickness dependence of the ISHE voltage in a YIG/Pt system.\cite{Castel2012} In contrast, the ANE-induced thermoelectric voltage in the PM configuration shows a monotonic increase of magnitude as the Pt thickness decreases [inset of Fig. \ref{fig2}(b)].\\
The observed ANE can originate from: the ANE of Fe$_3$O$_4$,\cite{Ramos2014a} or the presence of a magnetic interlayer at the Fe$_3$O$_4$/Pt interface. Under the first scenario the observed ANE voltage is expected to be strongly suppressed below the metal-insulator transition [Verwey transition, $T_\text{V} \sim$ 110 K for these films, see inset of Fig. \ref{fig2}(c)]. This is due to the even larger resistivity of Fe$_3$O$_4$ and increased shunting effect by the Pt layer. However, the temperature dependence of the thermoelectric voltage in the PM configuration [Fig. \ref{fig2}(c)] shows a non-negligible voltage even at temperatures lower than $T_V$, suggesting that the ANE signal might be originated at the Fe$_3$O$_4$/Pt interface.\\
Let us now focus on the thickness dependence of the ANE voltage measured in the PM configuration at room temperature. To analyze the result, we consider an equivalent circuit model in which a magnetic interlayer of thickness $t_\text{MP}$ is included at the Fe$_3$O$_4$/Pt interfaces [Fig. \ref{fig3}(a)]. This model describes the two previous scenarios where the ANE can originate from the Fe$_3$O$_4$ film ($E_\text{F}$) and at the interface region ($E_\text{MP}$). The resultant electric field in the non-magnetic Pt layer ($E_y$) can be expressed as:
  \begin{equation}
\label{eq:E_MP_thickness_dep}
  E_y = \frac{1}{1+\frac{\rho_\text{F}\rho_\text{MP}(t_\text{Pt}-2t_\text{MP})}{2\rho_\text{Pt}(t_\text{F}\rho_\text{MP}+t_\text{MP}\rho_\text{F})}}\left [\frac{t_\text{F}\rho_\text{MP}E_\text{F}+t_\text{MP}\rho_\text{F}E_\text{MP}}{t_\text{F}\rho_\text{MP}+t_\text{MP}\rho_\text{F}} \right ],
\end{equation}
where $\rho_\text{F}$ ($\rho_\text{MP}$, $\rho_\text{Pt}$) and $t_\text{F}$ ($t_\text{MP}$, $t_\text{Pt}$) represent the resistivity and thickness of Fe$_3$O$_4$ (MPE, Pt) layer, respectively. We consider $\rho_\text{F} = 6.96\times10^{-5}$ $\Omega$m,\cite{Ramos2015} and $\rho_\text{MP} = \rho_\text{Pt}$ (since they represent the magnetic and non-magnetic parts of the Pt layer). The thickness dependence of the Pt resistivity in the trilayers [Fig. \ref{fig3}(b)], is well described by conventional electron transport in metals:\cite{fuchs_1938,Sondheimer} $\rho_\text{Pt}=\rho_\infty \left (1 + \frac{3}{8(t_\text{Pt}-h)}(l_\infty(1-p)) \right )$, obtaining the fitting parameters: $\rho_{\infty}$ = 1.5 $\pm$ 0.2 $\times$ 10$^{-7}$ $\Omega$m, $h$ = 1.6 $\pm$ 0.7 nm and $l_\infty$ = 18 $\pm$ 9 nm, where p = 0 is assumed.\cite{Althammer2013} These are used in Eq. \ref{eq:E_MP_thickness_dep} to include the thickness dependence of $\rho_\text{Pt}$ into our model.\\
In Eq. \ref{eq:E_MP_thickness_dep} all the parameters are known, except for $E_\text{MP}$ and $t_\text{MP}$.
We can estimate the value of $E_\text{MP}$ using the model proposed by Guo et al.\cite{Guo2014}, where they theroretically estimated the ANE in a magnetized Pt as a function of the induced spin magnetic moment ($m_\text{S}$). With their estimation and the value of $m_\text{S}$ = 0.31 $\pm$ 0.04 $\mu_B$ for Pt recently measured by XMCD in a Fe$_3$O$_4$/Pt/Fe$_3$O$_4$ trilayer by Kikkawa et al.\cite{Kikkawa2017}, we obtain the induced ANE electric field as a function of the Pt resistivity ($E_\text{MP}/\nabla T$ = 2.5 $\rho_\text{Pt}$, where the resistivity prefactor units are Am$^{-1}$K$^{-1}$).\\
Now we can analyze the thickness dependence of the ANE using Eq. \ref{eq:E_MP_thickness_dep}. First, we evaluate the effect of the ANE of only the Fe$_3$O$_4$ layers, described by considering $t_\text{MP}$ = 0 nm as shown in Fig. \ref{fig3}(d), the magnitude of the $E_y$ thus obtained cannot explain our results, with a lower magnitude of the ANE. If we now introduce a non-negligible $t_\text{MP}$ we can reproduce the results with a value of the magnetic Pt thickness of about $t_\text{MP}$ = 0.1 nm. This thickness is significantly smaller than that obtained from XMCD and X-ray resonant magnetic reflectivity (XRMR) studies of Fe/Pt layers.\cite{Antel1999, Kuschel2015, Klewe2016} It is also smaller than the thickness of a possible interdiffusion region at the Fe$_3$O$_4$/Pt interface, as suggested by STEM elemental mapping measured by electron energy loss (EELS) and electron dispersive x-ray (EDX) spectroscopies in a [Fe$_3$O$_4$/Pt/Fe$_3$O$_4$] trilayer with $t_\text{Pt}$ = 10 nm [Fig. \ref{fig3}(c)], which suggest a small interface region of about 0.5 nm, where the Fe and Pt signals overlap. Even after considering a possible magnetic interdiffussion layer into our model, a non-negligible $t_\text{MP}$ is needed to explain the results (Suppl. Mat.\cite{SupplMat}). One possible scenario is that a subnanometer interdiffussion of Fe due to heating induces a modification of the Fe coordination and increases the Fe concentration at the interface.\cite{Willinger2014, Goniakowsk2009} This can result in an extrinsically induced Pt magnetic moment around Fe due to a modification of the nature of the F/N interface. This scenario is in agreement with a recent report,\cite{Vasili2018} and further supported by ANE measuremens of a Fe$_3$O$_4$/Pt bilayer, with Pt films deposited at two different temperatures (room temperature and 480 $^\circ$C, Suppl. Mat.\cite{SupplMat}), which show an increase of the ANE for Pt grown at higher temperature.\\
Finally, we would like to expand our model to describe the dependence of the interface-induced ANE on the number of multilayers. For a [Fe$_3$O$_4$/Pt]$_\text{n}$ multilayer, the expression of the ANE-induced electric field is:
\begin{widetext}
\begin{equation}
\label{eq:E_MP_ML}
  E_y = \frac{1}{1+\frac{\rho_\text{MP}\rho_\text{F}[nt_\text{Pt}-(2n-1)t_\text{MP}]}{\rho_\text{Pt}[nt_\text{F}\rho_\text{MP}+(2n-1)t_\text{MP}\rho_\text{F}]}}\left [\frac{nt_\text{F}\rho_\text{MP}E_\text{F}+(2n-1)t_\text{MP}\rho_\text{F}E_\text{MP}}{nt_\text{F}\rho_\text{MP}+(2n-1)t_\text{MP}\rho_\text{F}} \right ].
\end{equation}
\end{widetext}
Using the values of $E_\text{MP}$ and $t_\text{MP}$ previously estimated, we can obtain the dependence of the interface-induced ANE as a function of the number of layers  n. Fig. \ref{fig4} shows the comparison between the ANE measured in the PM configuration for two sets of multilayers with $t_\text{Pt}$ = 7 and 17 nm and the estimation using Eq. \ref{eq:E_MP_ML}. Our model can  describe the dependence of the ANE-induced electric field with the number of bilayers (n) and the layer thicknesses, the magnitude of the predicted ANE in the multilayers is slightly smaller than the measured one, specially in the case of thinner samples. The observed magnitude difference could be possibly due to the presence of additional contributions, such as interface-roughness-induced spin-orbit effects in the ferromagnetic layer,\cite{Zhou2015} which can generate ANE-like voltages as suggested in Ref. 18\nocite{Uchida2015a}.\\
In summary, we observed a non-negligible ANE in [Fe$_3$O$_4$/Pt]$_n$  multilayers and investigated its origin by systematic measurements in [Fe$_3$O$_4$/Pt/Fe$_3$O$_4$] trilayers as a function of the Pt thickness. The results can be understood by an interface-induced ANE, with its origin possibly due to a subnanometer Fe-Pt interdiffusion, which possibly affects the Fe coordination and/or the elemental composition at the interface. Measurements of magnetic moment and length-scale of magnetic interface ($t_\text{MP}$) as a function of Pt deposition temperature by other techniques (XMCD, XRMR) can help to further clarify the origin of the observed effect. These results suggest the possibility of tuning the thermoelectric response by thermal treatment and that, although the SSE is the dominant mechanism, the interface-induced ANE can positively contribute to the the spin-induced thermopower in multilayer systems.\\
\\
We thank T. Kuschel for fruitful discussions. This work was supported by ERATO ``Spin Quantum Rectification Project'' (Grant No. JPMJER1402) and PRESTO ``Phase Interfaces for Highly Efficient Energy Utilization'' (Grant No. JPMJPR12C1) from JST, Japan; Grant-in-Aid for Scientific Research (A) (Grant No. JP15H02012) and  Grant-in-Aid for Scientific Research on Innovative Area, ``Nano Spin Conversion Science'' (Grant No. JP26103005) from JSPS KAKENHI, Japan, the NEC Corporation and the Noguchi Institute. H2020-MSCA-RISE-2016 SPICOLOST (Grant No. 734187); the Spanish Ministry of Economy and Competitiveness (Grant No. MAT2017-82970-C2, including FEDER), Spain; and the Aragon regional government (E26), Spain. The microscopy works were conducted in the LMA at INA, Universidad de Zaragoza.

\bibliographystyle{apsrev}

\bibliography{bibliography}

\clearpage

\clearpage

\clearpage

\clearpage

\end{document}